\documentclass[twocolumn,showpacs,preprintnumbers,amsmath,amssymb,prb,aps]{revtex4-1}
\usepackage{epsfig}
\usepackage{epstopdf}
\usepackage{graphicx}
\usepackage{dcolumn}
\usepackage{bm}
\usepackage{slashbox}
\usepackage{textcomp}
\usepackage{array}
\usepackage{subcaption}
\usepackage{booktabs}

\begin{document}

\title{Thermoelectric properties of Fe$_{2}$VAl at high temperature region: A combined experimental and theoretical study}

\author{Shamim Sk$^{1,}$}
\altaffiliation{Electronic mail: shamimsk20@gmail.com}
\author{P. Devi$^{2,3}$}
\author{Sanjay Singh$^{2,4}$}
\author{Sudhir K. Pandey$^{5}$}
\altaffiliation{Electronic mail: sudhir@iitmandi.ac.in}
\affiliation{$^{1}$School of Basic Sciences, Indian Institute of Technology Mandi, Kamand - 175075, India}
\affiliation{$^{2}$Max Planck Institute for Chemical Physics of Solids, N\"{o}thnitzer Str. 40, 01187 Dresden, Germany}
\affiliation{$^{3}$Dresden High Magnetic Field Laboratory, Helmholtz-Zentrum Dresden-Rossendorf, Bautzner Landstr. 400, 01328 Dresden, Germany}
\affiliation{$^{4}$School of Materials Science and Technology, Indian Institute of Technology (Banaras Hindu University), Varanasi-221005, India}
\affiliation{$^{5}$School of Engineering, Indian Institute of Technology Mandi, Kamand - 175075, India}


\begin{abstract}
Heusler type compounds have long been recognized as potential thermoelectric (TE) materials. Here, the experimentally observed TE properties of Fe$_{2}$VAl are understood through electronic structure calculations in the temperature range of $300-800$ K. The observed value of \textit{S} is $\sim-$138 $\mu$V/K at 300 K. Then, the $|S|$ decreases with increase in temperature up to the highest temperature with the value of $\sim-$18 $\mu$V/K at 800 K. The negative sign of \textit{S} in the full temperature window signifies the dominating \textit{n}-type character of the compound. The temperature dependent of electrical conductivity, $\sigma$ (thermal conductivity, $\kappa$) exhibits the increasing (decreasing) trend with the values of $\sim$1.2 $\times$ 10$^{5}$ $\Omega^{-1}$m$^{-1}$ ($\sim$23.7 W/m-K) and $\sim$2.2 $\times$ 10$^{5}$ $\Omega^{-1}$m$^{-1}$ ($\sim$15.3 W/m-K) at 300 K and 800 K, respectively. In order to understand these transport properties, the DFT based semi-classical Boltzmann theory is used. The contributions of multi-band electron and hole pockets are found to be mainly responsible for the temperature dependent trend of these properties. The decrement of $|S|$ and increment of $\sigma/\tau$ $\&$ $\kappa_{e}/\tau$ ($\tau$ is relaxation time) with temperature is directly related with the contribution of multiple hole pockets. Present study suggests that DFT based electronic calculations provide reasonably good explanations of experimental TE properties of Fe$_{2}$VAl in the high temperature range of $300-800$ K.

\vspace{0.3cm}

\end{abstract}

\maketitle
\section{INTRODUCTION}
Fast consumption of natural energy sources pushes us to search for alternative sources of energy. In this context, Thermoelectric (TE) materials have gained much attention for addressing energy issues. TE material can convert the waste heat into useful electricity\cite{te1,te2}. But, the exploitation of TE materials is still suffered by their low efficiencies. The efficiency of TE material can be evaluated by its \textit{figure-of-merit} \textit{ZT}\cite{zt}, defined as $ZT=S^{2}\sigma T/(\kappa_{e}+\kappa_{L})$. Where, \textit{S} is the Seebeck coefficient, $\sigma$ the electrical conductivity and \textit{T} the absolute temperature of the sample. $\kappa_{e}$ and $\kappa_{L}$ are the electronic and lattice contributions to total thermal conductivity ($\kappa$). Efficient TE material should have the \textit{ZT} value greater than unity\cite{snyder}. Searching the high \textit{ZT} materials remains a challenging task, though the research in thermoelectrics has been going on for many decades. The value of \textit{ZT} can be improved by maximizing the power factor ($S^{2}\sigma$) or by minimizing the $\kappa$. Achieving high \textit{ZT} is really a challenging job as \textit{S}, $\sigma$ and $\kappa_{e}$ are strongly interrelated to each other through charge carriers\cite{ashcroft,shamim_mrx}. Hence, the systematic study of all these TE parameters at high temperature using experimental and theoretical approaches is required to understand the scenario in the better way. 

Heusler type compounds are a class of potential TE materials. The Heusler type structure is described by the formula $XYZ$ (half-Heuslers) or $X_{2}YZ$ (full-Heuslers), where \textit{X} and \textit{Y} are transition metals and \textit{Z} is in the \textit{p}-block element. These compounds possess face-centered cubic structure of $L2_{1}$ phase with space group \textit{Fm-3m}\cite{heusler1,heusler2,heusler3}. These compounds exhibit versatile properties and hence applications in many fields including spintronics\cite{heusler4,heusler5,heusler6}, topological insulators\cite{heusler7} etc. In the field of thermoelectricity, the Heusler compounds are found as potential candidates for the conversion of waste heat into useful electricity. These compounds are desirable for thermoelectric applications due to possessing large \textit{S}, moderate $\sigma$, high temperature stability, non-toxic, mechanical strength etc\cite{graf,zhu}. Number of experimental and theoretical TE studies on Heusler compounds have been carried out in last few decades\cite{sonu_jpcm,sonu_jpd,shastri_1st,shastri_3rd,lue_2002,lue_2004,muta,berry,shastri_4th,shastri_5th}. Among the existing Heusler compounds, the Fe$_{2}$VAl-based compounds have attracted considerable attention in the TE community\cite{alleno,anand}. Power factor of Fe$_{2}$VAl-based compounds are comparable with that of the state-of-the-art TE materials. But, the \textit{ZT} of this compound is diminished by its high $\kappa$. In order to inrease the \textit{ZT}, a rigorous efforts have been made by many groups via heavy element substitution\cite{nishino_2006,mikami,terazawa,lue_2008}, making of thin-film\cite{furuta} etc. But, still \textit{ZT} is smaller than 0.3 which is quite lower than the target value. Therefore, a systematic theoretical study is required to understand the experimental TE properties of Fe$_{2}$VAl in order to improve the \textit{ZT}.

For the calculations of transport properties, the density functional theory (DFT) based methods were found as more reliable and cheap among the available tools. In this method, the semi-classical Boltzmann transport equation has been solved by taking the ground state of DFT. Then, the required transport coefficients are calculated at finite temperatures. The temperature dependency of transport properties has been taken care of by Fermi-Dirac distribution function. In DFT calculation, the choosing of exchange-correlation (XC) functionals is one of the major challenges. The work of Sk \textit{et al}\cite{shamim_mrx} explores the scenario for choosing the XC functionals to understand the TE properties of Fe$_{2}$VAl. They showed that band gap of mBJ\cite{mbj} with the band-structure of PBEsol\cite{pbesol} or SCAN\cite{scan} gives the better explanation of experimental \textit{S} up to 620 K. Therefore, motivated by their work, here we have taken band gap of mBJ with band-structure of PBEsol for electronic structure and related transport properties calculations of Fe$_{2}$VAl up to 800 K. 

Keeping the above aspects in mind, here we try to understand the TE properties of Fe$_{2}$VAl through experimental and theoretical approaches in the temperature range of $300-800$ K. All the TE parameters S, $\sigma$ and $\kappa$ are measured using home-made instrumental setup. The room temperature value of \textit{S} is found to be $\sim-$138 $\mu$V/K, then it reaches $\sim-$18 $\mu$V/K at 800 K. The magnitude of \textit{S} decreases with increase in temperature in the full temperature range. The negative sign of \textit{S} throughout the studied temperature range signifies the dominating \textit{n}-type behaviour of the compound. The electrical conductivity shows the increasing nature with temperature, while thermal conductivity exhibits decreasing behaviour. In order to understand the experimental transport properties, we have carried out the electronic structure calculation in the framework of DFT. Semi-classical Boltzmann theory is used for transport calculations. Multi-band electron and hole pockets are found as mainly responsible for the temperature dependent behaviour of transport properties.

\section{EXPERIMENTAL AND COMPUTATIONAL DETAILS}
The polycrystalline Fe$_{2}$VAl was synthesized by arc melting technique using the appropriate amounts of Fe, V, and Al of 99.99\,\% purity. Then, the synthesized ingots were annealed at 1073 K for 5 days. Subsequently, the obtained compound was quenched into ice water to maintain the homogeneity. The room temperature x-ray diffraction confirms the $L2_{1}$ phase of face-centered cubic structure with lattice parameters 5.76 \AA, consistent with the literature\cite{boscow81}. The measurements of Seebeck coefficient and thermal conductivity were performed in the temperature range of $300-800$ K using home-made experimental setup\cite{shamim_instrument}. The resistivity was measured in the temperature range $300-720$ K using the in-house instrumental setup\cite{saurabh_resistivity}. The rectangular sample with dimension $\sim$ 5.0 mm $\times$ 3.8 mm $\times$ 1.3 mm was used for all the measurements.

The electronic structure calculations were done using a full-potential augmented plane wave (FP-LAPW) method as implemented in WIEN2k\cite{wien2k} code within density functional theory (DFT)\cite{dft}. The PBEsol\cite{pbesol} and mBJ\cite{mbj} were used as exchange-correlation (XC) functionals in the calculations. The lattice parameters of a = b = c = 5.76 \AA \, with spacegroup $Fm$-$3m$ were used for the calculations. The \textit{muffin-tin} sphere radii (R$_MT$) for Fe, V and Al were chosen as 2.35, 2.24 and 2.12 Bohr, respectively. The energy convergence criteria for self-consistent calculation was set to be 0.1 mRy/cell. The electronic transport properties were calculated using the BoltzTraP package\cite{boltztrap}. A heavy k-mesh of size 50 $\times$ 50 $\times$ 50 was used for the electronic structure calculations in order to get proper transport properties.

\begin{figure*}
\includegraphics[width=0.95\linewidth, height=5.2cm]{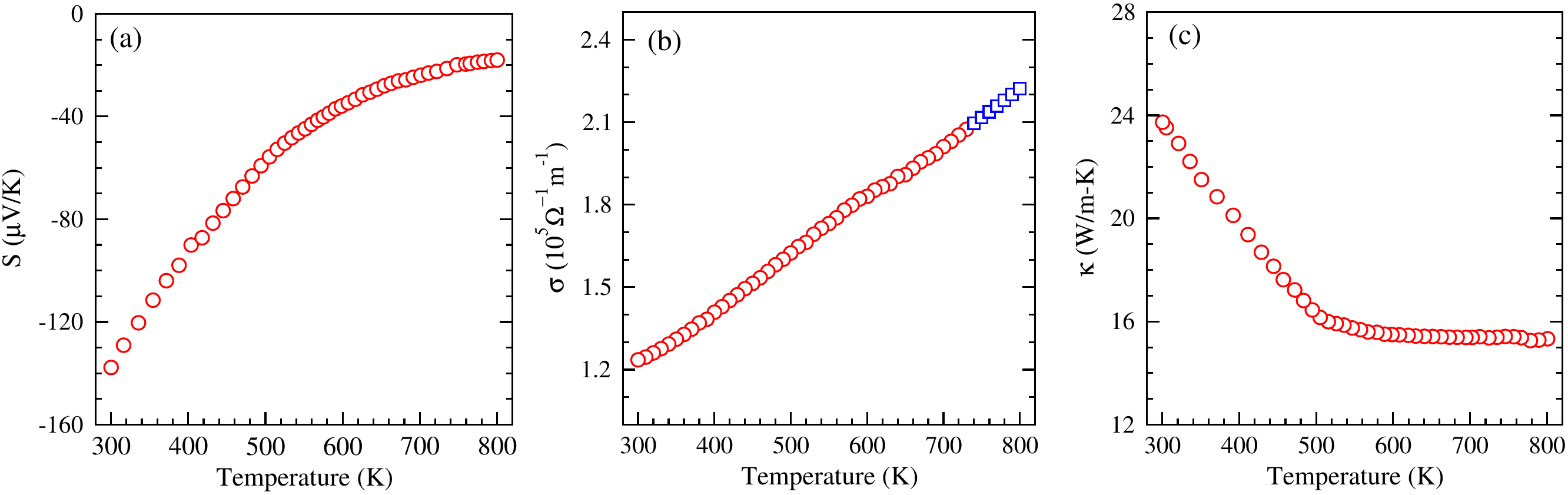} 
\caption{\small{Temperature dependence of (a) Seebeck coefficient, \textit{S}, (b) electrical conductivity, $\sigma$ and (c) thermal conductivity, $\kappa$.}}
\end{figure*}

\section{RESULTS AND DISCUSSION}        
\subsection{EXPERIMENTAL TRANSPORT PROPERTIES}     
Fig. 1(a) shows the temperature dependent \textit{S} of Fe$_{2}$VAl. At room temperature the \textit{S} is found to be $\sim-$138 $\mu$V/K. From the figure, it is noticed that with increase in temperature the $|S|$ decreases up to the highest temperature. But, the decrement rate of $|S|$ at high temperature region (above $\sim$500 K) becomes slower as compared to low temperature range. At 800 K, the observed value of \textit{S} is $\sim-$18 $\mu$V/K. Our result of temperature dependent \textit{S} is similar with the other reported works\cite{shamim_mrx,kurosaki}. The total \textit{S} in any material simply comes from the contributions of positively charged holes and negatively charged electrons. Under the framework of a two-carrier conduction model, the \textit{S} is defined as:\cite{ioffe} $S = \frac{S_{h}\sigma_{h}+S_{e}\sigma_{e}}{\sigma_{h}+\sigma_{e}}$, where $S_{h,e}$ and $\sigma_{h,e}$ represent the Seebeck coefficients and electrical conductivities for electrons and holes, respectively. The electrons usually give the negative \textit{S}, while holes give positive \textit{S}. Therefore, the negative sign of \textit{S} under the studied temperature range signifies the dominating \textit{n}-type behaviour of the compound. Specifically, at low temperature region the large magnitude of \textit{S} with negative sign reflects the large contribution of charge carriers from the electron pockets. As the temperature increases the contributions of charge carriers from electron and hole pockets are being compensated and hence gives the low \textit{S}. This behaviour of \textit{S} will be more clear when we will discuss the electronic band-structure to address the experimental \textit{S}.

Fig. 1(b) exhibits the measured electrical conductivity ($\sigma$) of Fe$_{2}$VAl in the temperature range of $300-800$ K. At 300 K, the $\sigma$ is found to be $\sim$1.2 $\times$ 10$^{5}$ $\Omega^{-1}$m$^{-1}$. The measurement of $\sigma$ was taken up to 730 K using the home-made instrumental setup. Then remaining data points up to 800 K are extrapolated (denoted by blue square symbol in the figure). The $\sigma$ at 800 K is observed as $\sim$2.2 $\times$ 10$^{5}$ $\Omega^{-1}$m$^{-1}$. The $\sigma$ is monotonically increasing in the full temperature range, consistent with the other reported work\cite{kurosaki}. The behaviour of $\sigma$ can be understood by using the well known formula: $\sigma = \frac{ne^{2}\tau}{m^{*}}$. Where, $n$ is the carrier concentration, $e$ is the electronic charge, $\tau$ is the relaxation time and $m^{*}$ is the effective mass of charge carriers. As the temperature increases $n$ always increases, while $\tau$ decreases for any material. Therefore, the dominating trend of $n$ or $\tau$ will give the temperature dependent feature of $\sigma$. Typically, in the case of semiconductor, the increment rate of $n$ is more dominant as compared to decrement rate of $\tau$. Therefore, temperature dependent $\sigma$ of Fe$_{2}$VAl is showing the semiconducting like behavior in the entire temperature range. 

The measurement of thermal conductivity ($\kappa$) for Fe$_{2}$VAl is shown in Fig. 1(c) in the temperature range of $300-800$ K. The value of $\kappa$ at room temperature is found to be $\sim$23.7 W/m-K. The $\kappa$ is decreasing up to the highest temperature. From the figure, it is clear that the $\kappa$ is decreasing rapidly up to $\sim$530 K with the value of $\sim$15.9 W/m-K. After 530 K, the decrement of $\kappa$ is very slow till the highest temperature with the value of $\sim$15.3 W/m-K at 800 K. Our result is also consistent with the other reported work\cite{kurosaki}. The average decrement rate in the temperature range $300-530$ K is calculated as $\sim$0.03 W/m-K, whereas this value reaches $\sim$0.002 W/m-K in the temperature range $530-800$ K. The values of $\kappa$ for Fe$_{2}$VAl are relatively high as compared to so called state-of-the-art TE materials. The value of \textit{ZT} for Fe$_{2}$VAl is mainly suppressed by this high value of $\kappa$. The $\kappa$ consists of two parts: electronic thermal conductivity ($\kappa_{e}$) and lattice thermal conductivity ($\kappa_{L}$). The $\kappa_{e}/\tau$ is calculated in this work and described later.

\subsection{ELECTRONIC STRUCTURE}
To understand the experimentally observed transport properties, the electronic structure calculations have been carried out. Fig. 2(a) shows the calculated band-structure of Fe$_{2}$VAl using PBEsol functional along the high symmetry directions $W-L-\Gamma-X-W-K$ in the first Brillouin zone. The four bands indexing by numbers 1, 2, 3 and 4 around the Fermi level (E$_{F}$) are expected to contribute to the transport properties. From the figure, it is seen that few of the occupied bands (bands 1, 2 and 3) and unoccupied band (band 3) are crossing the E$_{F}$ at $\Gamma$ and $X$ points, respectively. This type of band-feature generally predicts the semi-metal like behaviour of the compound. Figure also shows that top of the occupied bands at $\Gamma$-point are triply degenerated (bands 1, 2 and 3), while bottom of the unoccupied band at $X$-point is non-degenerated (band 4). At this point, it is important to mention that our previous study suggests that the band-feature of PBEsol with the band gap of mBJ is the good choice to address the experimental \textit{S}\cite{shamim_mrx}. Subsequently, we have given the rigid band-shift to the band-feature of PBEsol in order to get the mBJ band gap of 0.22 eV, which is shown in Fig. 2(b). The red dashed line indicates the E$_{F}$, which is set at the middle of the band gap. Here, the features of Fig. 2(a) and 2(b) are same, but only difference in the rigid band-shift. However, the Fig. 2(b) has been used to calculate the electronic transport properties in this work. Therefore, we will come back in this figure once we discuss the calculated transport properties in the next sub-section.

\begin{figure*} 
\begin{subfigure}{0.245\textwidth}
\includegraphics[width=0.99\linewidth, height=3.8cm]{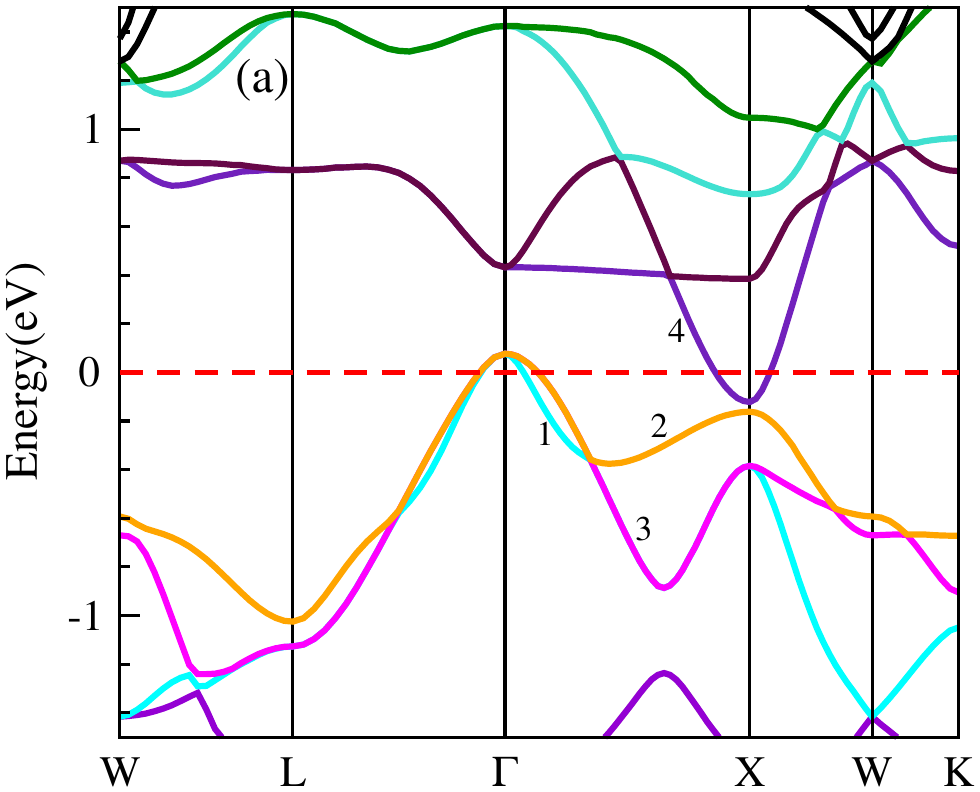} 
\end{subfigure}
\begin{subfigure}{0.245\textwidth}
\includegraphics[width=0.99\linewidth, height=3.8cm]{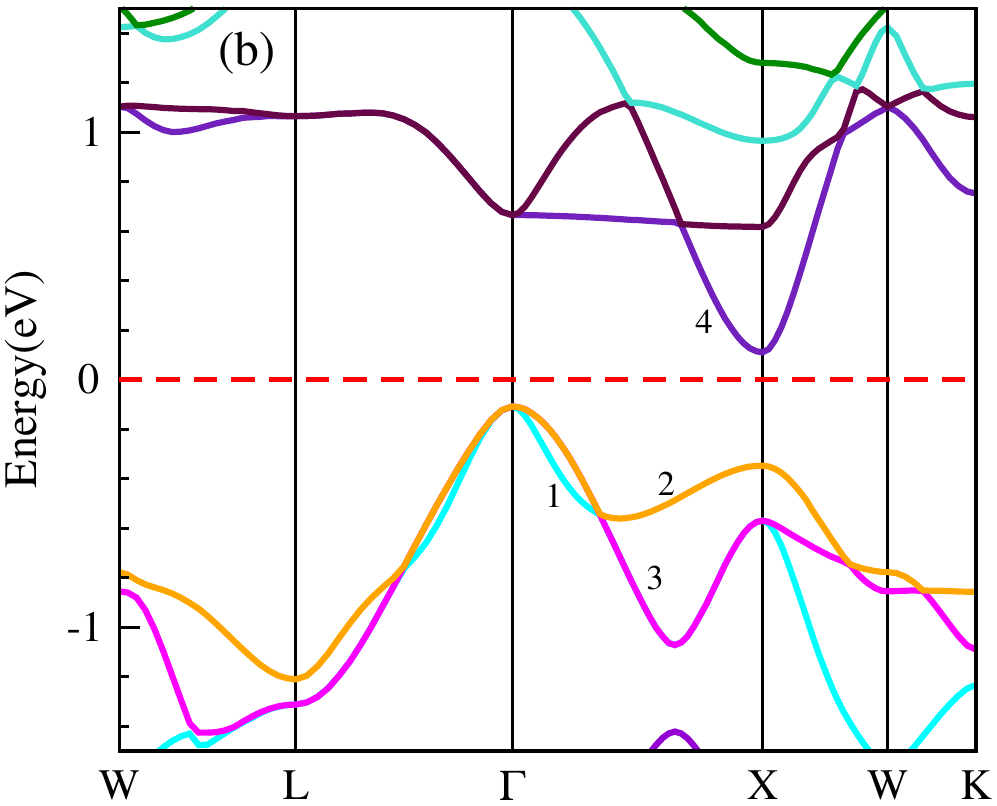} 
\end{subfigure}
\begin{subfigure}{0.49\textwidth}
\includegraphics[width=0.99\linewidth, height=4.0cm]{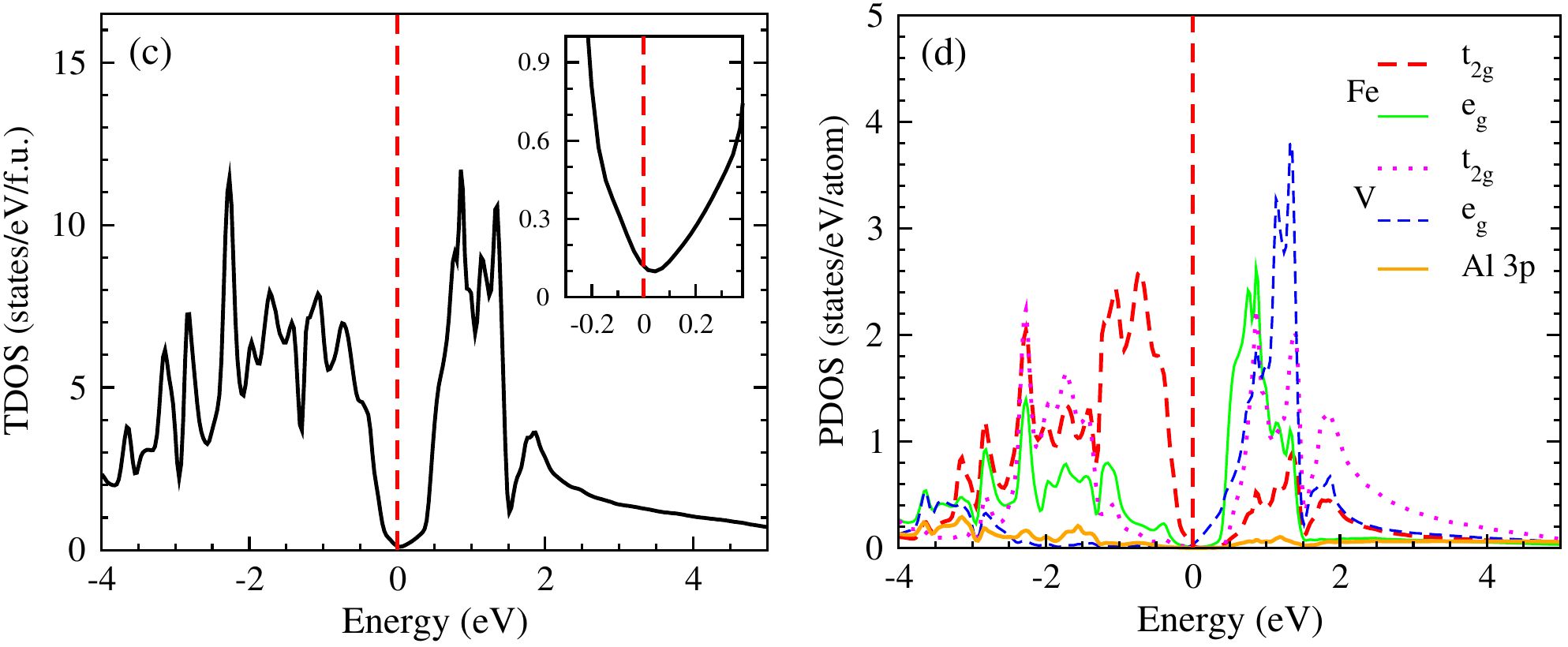}
\end{subfigure} 
\caption{\small{(a) The calculated band-structure using PBEsol. Red dashed line indicates the Fermi level. (b) Band-structure with rigid band shift. Where, band gap is taken from mBJ. (c) Total density of states (TDOS). (d) Partial density of states (PDOS).}}
\label{fig:image2}
\end{figure*}

Fig. 2(c) presents the calculated total density of states (TDOS) for Fe$_{2}$VAl. Minimum DOS around the E$_{F}$ (see the inset of Fig. 2(c)) signifies the presence of pseudogap, which predicts the semi-metal like behaviour of the compound. To see the contribution in transport properties from different atomic orbitals, we have calculated the partial density of states (PDOS) for all the constituent atoms as shown in Fig. 2(d). Here, we have only shown the PDOS of Fe 3\textit{d}, V 3\textit{d} and Al 3\textit{p} orbitals, which give the main contribution in DOS around E$_{F}$. The five degenerate \textit{d} orbitals are shown as three-fold degenerate states $t_{2g}$ ($d_{xy}$, $d_{yz}$, $d_{zx}$) and two-fold degenerate states $e_{g}$ ($d_{x^{2}-y^{2}}$, $d_{z^{2}}$) in the figure. From the figure, it is clear that the participation of Al 3\textit{p} orbitals in DOS around the E$_{F}$ is negligibly small as compared to Fe 3\textit{d} and V 3\textit{d} orbitals. In the low-lying energy range of $\sim-$1.0 to 0 eV, the dominant contribution in DOS comes from Fe $t_{2g}$ orbitals with small contribution from Fe $e_{g}$ and V $t_{2g}$. Above E$_{F}$, the Fe $e_{g}$, V $t_{2g}$ and V $e_{g}$ orbitals are mainly participating in DOS with small contribution from Fe $t_{2g}$. 

In order to understand the effect of Effective mass ($m^{*}$) on transport properties, we have calculated the $m^{*}$ of electrons and holes at the vicinity of high symmetric k-points in terms of electrons mass ($m_{e}$). Fig. 2(b) shows that the bands 1, 2, 3 and 4 are significantly contributed to the transport properties. Therefore, the $m^{*}$ of these bands named as B1, B2, B3 and B4 are calculated at the vicinity of $\Gamma$ and $X$-points, which are tabulated in Table I. The $m^{*}$ is calculated using the formula: $m^{*}=\hbar^{2}/(d^{2}E/dk^{2})$ under parabolic approximation\cite{ashcroft}. Formula suggests that a flat band will give a large $m^{*}$ as compared to a curved band. The electron pocket (concave band) gives the positive $m^{*}$, while hole pocket (convex band) gives the negative $m^{*}$. Table I shows only the magnitude of $m^{*}$ of charge carriers. The calculations of $m^{*}$ will be helpful to explain the transport properties in the next sub-section.

\begin{table}
\caption{\small{Calculated effective mass ($m^{*}$) at the vicinity of $\Gamma$ and \textit{X}-points. }}
\resizebox{0.45\textwidth}{!}{%
\setlength{\tabcolsep}{6pt}
\begin{tabular}{@{\extracolsep{\fill}}c c c c c c c c c c c} 
\hline\hline
 
\multicolumn{1}{c}{$m^{*}$ at the} & & & &  \multicolumn{4}{c}{Bands}  &&  \multicolumn{1}{c}{} & \multicolumn{1}{c}{}\\ 
\cline{3-11}                                
  vicinity of & & \multicolumn{1}{c}{B1} &&   \multicolumn{1}{c}{B2}  &&   \multicolumn{1}{c}{B3}   &&   \multicolumn{1}{c}{B4}\\
   
 \hline
$\Gamma$-point && 0.51 &&  0.90  && 0.90  && $-$ \\
$X$-point && $-$ && 1.84  && $-$   &&  0.49 \\

\hline\hline
 
\end{tabular}}
\end{table} 

\subsection{CALCULATED TRANSPORT PROPERTIES}
Here, the temperature dependence of Seebeck coefficient (\textit{S}), electrical conductivity per relaxation time ($\sigma/\tau$) and electronic thermal conductivity per relaxation time ($\kappa_{e}/\tau$) are discussed. These electronic transport coefficients are calculated using the BoltzTraP package\cite{boltztrap}, where semi-classical Boltzmann theory has been implemented. The band-structure obtained from PBEsol with mBJ band gap (Fig. 2(b)) is used as a key input in the BoltzTraP package. First of all, the room temperature value of \textit{S} is calculated as $\sim$177 $\mu$V/K at E$_{F}$. This high value of \textit{S} with positive sign can be explained by the electron/hole pockets and effective mass ($m^{*}$) of charge carriers. The \textit{S} is directly proportional to $m^{*}$ of charge carriers\cite{snyder}. The energy gap from the E$_{F}$ to top of the occupied band (OB) at $\Gamma$-point and bottom of the unoccupied band (UOB) at $X$-point are same as seen from Fig. 2(b). Therefore, there will be equal probability to excite the holes from hole pocket (at $\Gamma$-point) and electrons from electron pockets (at $X$-point). But, the $m^{*}$ of holes corresponding to bands 2 and 3 (at the vicinity of $\Gamma$-point) are almost double as compared to $m^{*}$ of electron corresponding to band 4 (at the vicinity of $X$-point) as seen from Table I. That's why holes are more dominant in \textit{S} than electrons and we are getting high \textit{S} with positive sign at E$_{F}$. However, this value of \textit{S} is very much far away from the experimental value of $\sim-$138 $\mu$V/K. At this point, it is important to note that the calculation of \textit{S} is done on a stoichiometric sample. But, there is  always a chance to have the off-stoichiometry in any real sample. The possible sources for this off-stoichiometry are: inaccuracy in weighing the starting materials, inhomogeneous mixing during the sample synthesis, evaporation of low melting element during the heat treatment etc. All these factors can be attributed in the calculation by considering the off-stoichiometry of the sample through the shifting of chemical potential ($\mu$). We found at $\mu$ $\approx$ 20 meV, the calculated values of \textit{S} give a good explanation of experimental \textit{S}. Then at $\mu$ $\approx$ 20 meV, the \textit{S} is calculated in the temperature range of $300-800$ and compared with the experimental \textit{S} as shown in Fig. 3(a). Figure shows the good match of calculated \textit{S} with experiment. But, as the temperature increases, a little deviation between calculated and experimental \textit{S} is observed from the figure.

\begin{figure*}
\includegraphics[width=0.98\linewidth, height=5.5cm]{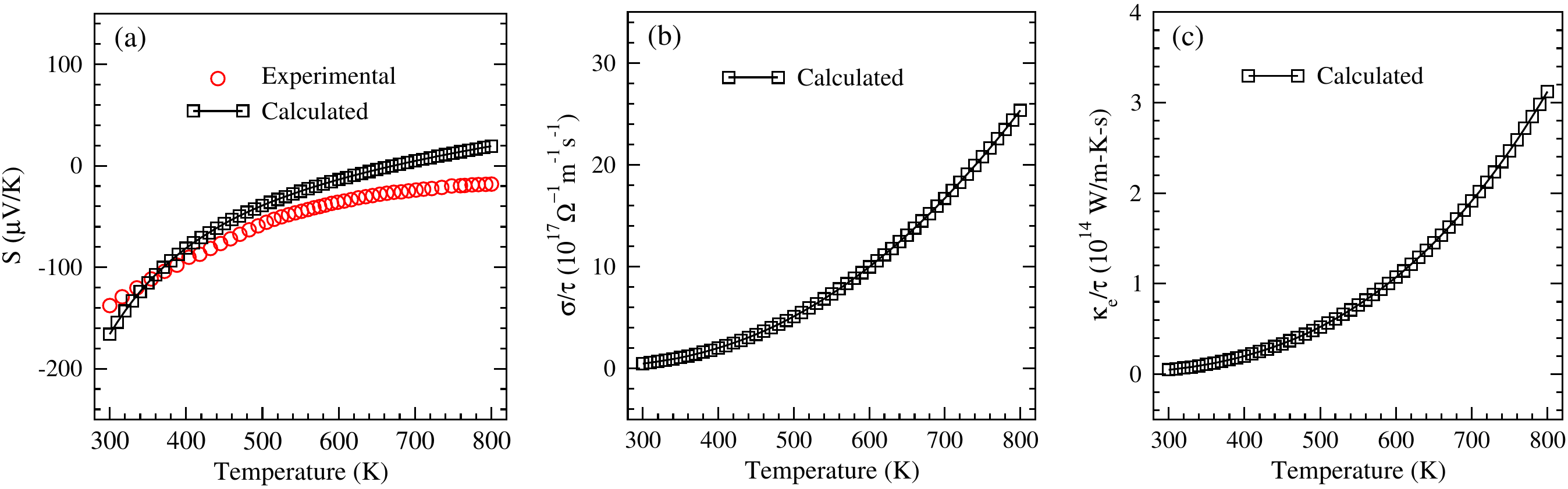} 
\caption{\small{(a) Comparison between calculated and experimental Seebeck coefficient, (b) Calculated electrical conductivity per relaxation time and (c) Calculated  thermal conductivity per relaxation time as a function of temperature.}}
\end{figure*}

Again the temperature dependent behaviour of calculated \textit{S} at $\mu$ $\approx$ 20 meV can be understood by the effective mass of charge carriers corresponding to hole (bands 1, 2, and 3) and electron (band 4) pockets at the vicinity of $\Gamma$ and $X$-points, respectively. Now, the $\mu$ is shifted towards the UOB and hence the energy gap between this $\mu$ and bottom of the UOB is effectively reduced. Therefore, electrons from the electron pocket (at $\Gamma$-point) will be easily excited than the holes from the hole pockets (at $X$-point). This is the reason why we are getting negative sign of \textit{S} with high magnitude at low temperature region. But, as the temperature increases, the magnitude of \textit{S} decreases, because at high temperature the holes from hole pockets jumps the energy gap and start to contribute to \textit{S}. The effective mass of hole pockets (mainly bands 2 and 3) are sufficiently larger than the effective mass of electron (at $\Gamma$-point). Hence, the further increment in temperature is expected to provide the positive sign of \textit{S}, where holes will be the majority contributors to \textit{S}. The same is observed from the Fig. 3(a), above $\sim$700 K, the S is positive. At the ultra-high temperature, the massive number of holes from the hole pocket at the vicinity of $X$-point (large $m^{*}$, band 2) will also contribute to the \textit{S} and high magnitude of positive \textit{S} is expected. Fig. 3(a) shows that as the temperature increases, the calculated values of \textit{S} are deviating from the experimental \textit{S}. At this point, it is important to note that the calculation of \textit{S} is carried out by taking the ground-state band-structure at constant $\mu$. But, in general the band-structure and $\mu$ are temperature dependent quantities. Therefore, the high temperature result of \textit{S} can be improved by considering the temperature dependent band-structure and $\mu$. However, for this study, once needs to go beyond the DFT method which demands more computational cost and also out of the scope of our present study. Here, it is important to mention that the same $\mu$ value of 20 meV is used to calculate the other electronic transport coefficients in this study.

The $\boldsymbol{\sigma}/\tau$ is calculated in the temperature range $300-800$ K as shown in Fig. 3(b). Here, also the electronic band-structure is used to calculate $\boldsymbol{\sigma}$. The total $\boldsymbol{\sigma}$ is computed by adding the $\boldsymbol{\sigma}$ of all bands as: $\boldsymbol{\sigma} = \sum_{n}\boldsymbol{\sigma}^{(n)}$. Where, $\boldsymbol{\sigma}$ of $n^{th}$ band is defined as\cite{ashcroft}:
\begin{equation}
\boldsymbol{\sigma}^{(n)} = e^{2}\int\frac{d\textbf{k}}{4\pi^{3}}\tau_{n}(\varepsilon_{n}(\textbf{k}))\textbf{v}_{n}(\textbf{k})\textbf{v}_{n}(\textbf{k})\bigg(-\frac{\partial f}{\partial \varepsilon}\bigg)_{\varepsilon=\varepsilon_{n}(\textbf{k})},
\end{equation}
where, $e$ is an electronic charge, $\tau_{n}(\varepsilon_{n}(\textbf{k}))$ and $\textbf{v}_{n}(\textbf{k})$ are relaxation time and mean velocity of an electron of $n^{th}$ band with wave vector \textbf{k}, respectively. The $\varepsilon_{n}(\textbf{k})$ is an energy band and $\frac{\partial f}{\partial \varepsilon}$ is the partial energy derivative of Fermi-Dirac distribution function. Here, the Fermi-Dirac distribution function is responsible for the temperature dependent study. At this point, it is important to note that the BoltzTraP package is based on the constant relaxation time approximation, i.e. $\tau_{n}(\varepsilon_{n}(\textbf{k}))$ is simply taken as $\tau$. Therefore, the final value of $\boldsymbol{\sigma}/\tau$ depends on the number of available states at a given $\mu$, $\textbf{v}_{n}(\textbf{k})$ and $\frac{\partial f}{\partial \varepsilon}$. Fig. 3(b) shows that the $\boldsymbol{\sigma}/\tau$ increases as the temperature increases. With increase in temperature more states will be available for the conduction and increment of $\boldsymbol{\sigma}/\tau$ is expected. If we observe the band-structure plot of Fig. 2(b), we can see that initially the charge carriers of band 4 from $\Gamma$-point are contributed to the $\boldsymbol{\sigma}/\tau$. But, as the temperature increases, the large number of charge carriers from bands 1, 2 and 3 participate to the conduction. Instantly, the $\boldsymbol{\sigma}/\tau$ increases as the temperature increases. The values of $\boldsymbol{\sigma}/\tau$ are found to be $\sim$4.9 $\times$ 10$^{16}$ and $\sim$2.5 $\times$ 10$^{18}$ $\Omega^{-1}$m$^{-1}$s$^{-1}$ at 300 and 800 K, respectively. The calculated values of $\boldsymbol{\sigma}$ can be compared with the experiment if temperature dependent $\tau$ is available. 

The $\kappa_{e}/\tau$ is also calculated using the same semi-classical Boltzmann theory. Fig. 3(c) exhibits the calculated values of $\kappa_{e}/\tau$ in the temperature range $300-800$ K. The value of $\kappa_{e}/\tau$ is found to be $\sim$0.05 $\times$ 10$^{14}$ W/m-K-s. Thereafter, as the temperature increases, the $\kappa_{e}/\tau$ also increases in the full temperature range. At 800 K, the calculated value of $\kappa_{e}/\tau$ is $\sim$3.12 $\times$ 10$^{14}$ W/m-K-s. The feature of temperature dependent $\kappa_{e}/\tau$ can be understood by the following equation\cite{ashcroft}
\begin{equation}
\boldsymbol{\kappa_{e}}/\tau=\frac{\pi^{2}}{3}\bigg(\frac{k_{B}}{e}\bigg)^{2}T(\boldsymbol{\sigma/\tau}),
\end{equation}            
where the Boltzmann constant, $k_{B}$ and electronic charge, $e$ are constants. Therefore, the trend of $\kappa_{e}/\tau$ solely depends on $T$ and $\boldsymbol{\sigma/\tau}$. We have seen that from Fig. 3(b), the values of $\boldsymbol{\sigma}/\tau$ increase with increase in temperature. Therefore, both the $T$ and $\boldsymbol{\sigma/\tau}$ are responsible for the increment nature of $\kappa_{e}/\tau$ in Fig. 3(c). For more clarification, we can see the electronic-dispersion of Fig. 2(b), where the more charge carriers are participated for the conduction of heat from bands 1, 2, and 3 (at the vicinity of $\Gamma$-point) as the temperature increases. Due to this, the increment behaviour of $\kappa_{e}/\tau$ is observed under the increment of temperature.

\subsection{FIGURE OF MERIT, \textit{ZT}}
The applicability of TE materials is decided by it's value of \textit{ZT}. Fig. 4 shows the experimental \textit{ZT} of Fe$_{2}$VAl in the temperature range of $300-800$ K. The value of \textit{ZT} is found to be $\sim$0.02 at 300 K. Then the value of \textit{ZT} decreases as the temperature increases up to the highest temperature studied here. At 800 K, the value of \textit{ZT} is 0.0025. This temperature dependent decrement nature of \textit{ZT} is due to the decreasing trend of $|S|$ with temperature. The observed value of \textit{ZT} is quite low as compared to so called state-of-the-art TE materials. Actually, the high $\kappa$ value of this compound is responsible for the low \textit{ZT}. Most of the contribution in total $\kappa$ comes from $\kappa_{L}$. Therefore, the massive amount of heat are transferred through lattice rather than electrons in this compound. Hence, a rigorous effort should be made to reduce $\kappa_{L}$ in order to get high \textit{ZT}. Alloying and nanostructuring are two of the many important techniques to reduce $\kappa_{L}$\cite{snyder,djsingh}. 

\begin{figure}
\includegraphics[width=0.75\linewidth, height=6.0cm]{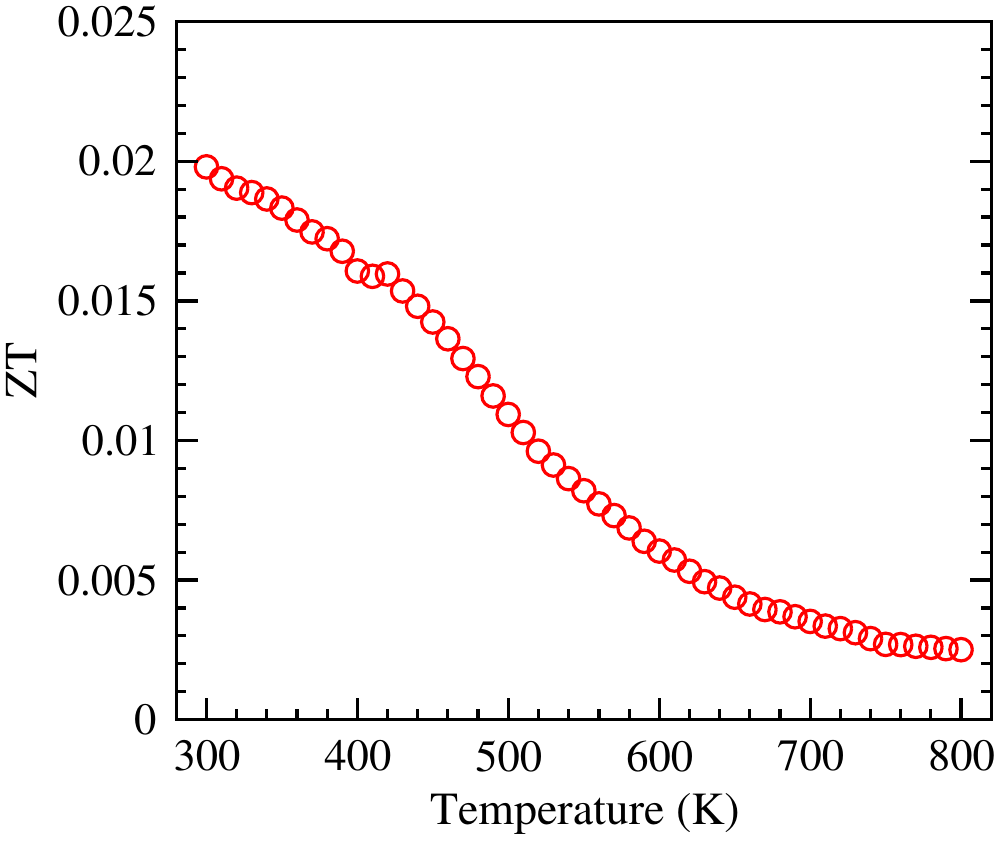} 
\caption{\small{Temperature dependence of \textit{figure-of-merit}, \textit{ZT}.}}
\end{figure}

\section{CONCLUSIONS}
In conclusion, we have analyzed the experimentally observed TE properties of Fe$_{2}$VAl using the electronic structure calculations in the temperature range of $300-800$ K. The values of \textit{S} are found to be $\sim-$138 $\mu$V/K and $\sim-$18 $\mu$V/K at 300 K and 800 K, respectively. The magnitude of \textit{S} decreases up to the highest studied temperature. But, the decrement rate of $|S|$ at the high temperature region (above $\sim$500 K) is lower than the low temperature range. In the full temperature window, \textit{S} exhibits the negative sign indicating the dominating \textit{n}-type character of the compound. The electrical conductivity ($\sigma$) shows the increasing trend with temperature, whereas thermal conductivity ($\kappa$) exhibits the decreasing trend in the full temperature range. The values of $\sigma$ ($\kappa$) are observed as $\sim$1.2 $\times$ 10$^{5}$ $\Omega^{-1}$m$^{-1}$ ($\sim$23.7 W/m-K) and $\sim$2.2 $\times$ 10$^{5}$ $\Omega^{-1}$m$^{-1}$ ($\sim$15.3 W/m-K) at 300 K and 800 K, respectively. The DFT based methods are used to understand these transport properties. The temperature dependent of these properties are explained by the contributions of multi-band electron and hole pockets from electronic dispersion. This study proposes that DFT based electronic calculations can be used to address the experimental TE properties of materials at high temperature. However, for more quantification in the results, one may go beyond DFT based methods which requires heavy computational cost. Therefore, the reliable and relatively cheap DFT based methods can be implemented for searching the new TE materials for high temperature applications.  \\          

\section{ACKNOWLEDGEMENTS}
S Singh acknowledges support from Science and Engineering Research Board of India for financial support through the Ramanujan Fellowship and Early Career Research Award.

\end{document}